\documentclass[aps,prl,twocolumn,superscriptaddress,showpacs,letter]{revtex4-1}
\usepackage{graphicx}
\usepackage{dcolumn}
\usepackage{bm}
\usepackage{cancel}
\usepackage{color}
\usepackage{amssymb}
\newcommand{\ba}{\begin{array}}
\newcommand{\ea}{\end{array}}

\def\br{\begin{eqnarray}}
\def\er{\end{eqnarray}}
\def\be{\begin{equation}}
\def\ee{\end{equation}}

\def\({\left(}
\def\){\right)}

\begin{document}
      
\title{Proton-proton forward scattering at the LHC}

\author{M.~Broilo}
\email{mateus.broilo@ufrgs.br}
\affiliation{Instituto de F\'isica e Matem\'atica, Universidade Federal de Pelotas, 96010-900, Pelotas, RS, Brazil}
\affiliation{Instituto de F\'isica, Universidade Federal do Rio Grande do Sul, Caixa Postal 15051, 91501-970, Porto Alegre, RS, Brazil}
\author{D.~A.~Fagundes}
\email{daniel.fagundes@ufsc.br}
\affiliation{Departamento de Ci\^encias Exatas e Educa\c{c}\~ao, Universidade Federal de Santa Catarina - Campus Blumenau, 89065-300, Blumenau,
SC, Brazil}
\author{E.~G.~S.~Luna}
\email{luna@if.ufrgs.br}
\affiliation{Instituto de F\'isica, Universidade Federal do Rio Grande do Sul, Caixa Postal 15051, 91501-970, Porto Alegre, RS, Brazil}
\affiliation{Instituto de F\'{\i}sica, Facultad de Ingenier\'{\i}a, Universidad de la Rep\'ublica \\
J.H. y Reissig 565, 11000 Montevideo, Uruguay}
\author{M.~J.~Menon}
\email{menon@ifi.unicamp.br}
\affiliation{Instituto de F\'{\i}sica Gleb Wataghin, Universidade Estadual de Campinas, 13083-859, Campinas, SP, Brazil}


\begin{abstract}

Recently the TOTEM experiment at the LHC has released measurements at $\sqrt{s} = 13$ TeV of the proton-proton total cross section,
$\sigma_{tot}$, and the ratio of the real to imaginary parts of the forward elastic amplitude, $\rho$. Since then an intense debate
on the $C$-parity asymptotic nature of the scattering amplitude was initiated. We examine the proton-proton and the antiproton-proton forward data
above 10 GeV in the context of an eikonal QCD-based model, where nonperturbative effects are readily included via a QCD effective charge. We show
that, despite an overall satisfactory description of the forward data is obtained by a model in which the scattering amplitude is dominated by only
crossing-even elastic terms, there is evidence that the introduction of a crossing-odd term may improve the agreement with the measurements of
$\rho$ at $\sqrt{s} = 13$ TeV.
In the Regge language the dominant even(odd)-under-crossing object is the so called Pomeron (Odderon).

\end{abstract}


\maketitle

\section{Introduction}

The rise of the total cross section with energy in hadron-hadron collisions was theoretically predicted many years ago
\cite{cheng001}.
This rise, as indicated by the data accumulated from collider and cosmic ray experiments over the last several decades, is dictated by jets with
transverse energy $E_{T}$ much smaller than the total energy $s$ available in the hadronic collision \cite{evidences001}. From the QCD viewpoint,
these {\it minijets}
arise from semihard scatterings of elementary partons, defined as hard scatterings of partons carrying very small fractions of momenta of their parent
hadrons \cite{ryskin001,ryskin002}. In this picture the high-energy behavior of the cross sections is driven mainly by semihard processes involving
gluons, since they give the dominant contribution at small $x$ \cite{qcdmodel001,qcdmodel002,luna001}. 
In accordance with very general predictions based on axiomatic field theory, all models using this QCD-based formalism have assumed over the years
that at high
energies the scattering amplitude is dominated by only a single crossing-even elastic amplitude. In the Regge language this dominant
even-under-crossing object
is named Pomeron \cite{barone001}; in the QCD language the Pomeron is, in its simplest configuration, a color singlet made up of two gluons
\cite{bfkl,forshaw001}.

Very recently the TOTEM experiment at the LHC has released measurements at $\sqrt{s} = 13$ TeV of the proton-proton ($pp$) total cross section,
$\sigma_{tot} = 110.6 \pm 3.4$ mb \cite{antchev001},
and in a subsequent work another independent measurement, $\sigma_{tot} = 110.3 \pm 3.5$ mb, together
with measurements of the ratio real-to-imaginary parts of the
forward amplitude,
$\rho = 0.09 \pm 0.01$ and $\rho = 0.10 \pm 0.01$ \cite{antchev002}.
These measurements can be simultaneously well-described by models that include the odd-under-crossing partner of
the Pomeron: The Odderon \cite{odderon001,odderon002}. The Odderon concept is physically reasonable since its presence does not violate asymptotic
theorems based on analyticity, unitarity and crossing. Also from the QCD viewpoint the Odderon is, in its simplest configuration, a color singlet made
up of three gluons \cite{ewerz001}. More specifically, in perturbative QCD the Odderon can be associated to a colorless $C$-odd $t$-channel state
with an intercept at or near one.
The TOTEM results have since then triggered an intense debate on the question of whether or not the combined behavior of $\sigma_{tot}$ and $\rho$ at
high energies is actually a manifestation of the Odderon \cite{odderon002,odderon003}.

This Letter revisits this issue by investigating the behavior of the forward quantities $\sigma_{tot}$ and $\rho$, from $pp$ and $\bar{p}p$
(antiproton-proton) scattering, through an eikonal
QCD-based model with a scattering amplitude dominated asymptotically by only single crossing-even amplitudes \cite{luna001,luna002,luna003}.
We consider a model in which the
bridge between the parton dynamics, described by QCD, and the dynamics of observable hadron systems is provided by the QCD parton model using updated
sets of quark and gluon distribution functions, standard QCD cross sections for parton-parton processes and physically motivated
cutoffs that restrict the parton-level dynamics to the semihard sector. The nonperturbative dynamics of the QCD is treated in the context of a
well-established infrared effective charge dependent on the dynamical gluon mass. 
Such a model, involving only even-under-crossing amplitudes dominant at very high energies, provides a satisfactory (from a statistical viewpoint)
global description of $\sigma_{tot}$ and $\rho$ data over a wide range of energies, and furthermore, the larger $\rho$ value predicted by the model
at $\sqrt{s}=13$ TeV suggests that a crossing-odd elastic term may play an important role in the soft and semihard interactions.

\section{The Model} 

The unitarity of the matrix $S$ requires that the absorptive part of the elastic scattering amplitude receives
contributions from both elastic and inelastic channels. In impact parameter ($b$) representation the unitarity condition implies the relation
\begin{eqnarray}
2 \textnormal{Re}\, h(s,b) = |h(s,b)|^{2} + G_{inel}(s,b) ,
\label{unitarity001}
\end{eqnarray}
where $h(s,b)$ is the profile function and $G_{inel}(s,b)$ is the inelastic overlap function, i.e. the contribution from all inelastic channels.
The profile function $h(s,b)$ is related to the elastic scattering amplitude ${\cal A}(s,t)$ by
\begin{eqnarray}
{\cal A}(s,t) = i \int_{0}^{\infty}b\, db\, J_{0}(b\sqrt{-t})\, h(s,b),
\label{scatter001}
\end{eqnarray}
where $t$ is the usual Mandelstam variable.
In order to satisfy unitarity constraints, the profile function is conveniently written as $h(s,b) = 1-e^{-\chi(s,b)} $, where the eikonal function
$\chi(s,b)$ is a complex generalized phase-shift. In terms of the eikonal function the elastic amplitude reads
\begin{eqnarray}
{\cal A}(s,t) = i \int_{0}^{\infty} b\, db\, J_{0}(b\sqrt{-t}) \left[1-e^{-\chi(s,b)} \right],
\label{equation02}
\end{eqnarray}
where $\chi(s,b)=\textnormal{Re}\, \chi(s,b) + i\, \textnormal{Im}\, \chi(s,b)\equiv\chi_{_{R}}(s,b)+i\chi_{_{I}}(s,b)$. Hence the scattering amplitude is
completely determined once the eikonal function is known. The eikonal is written in terms of even and odd parts connected by crossing symmetry. In the
case of $pp$ and $\bar{p}p$ channels, this combination reads $\chi^{\bar{p}p}_{pp}(s,b)=\chi^{+}(s,b)\pm\chi^{-}(s,b)$.
Two observables that play a key role to unravel the structure of the elastic scattering amplitude at high energies,
namely $\sigma_{tot}$ and $\rho$, can be written in terms of $\chi(s,b)$. We see from the optical theorem that the total cross section is such that
\begin{eqnarray}
\sigma_{tot}(s) = 4\pi\, \textnormal{Im}\,{\cal A}(s,t=0), 
\end{eqnarray}
thus, from (\ref{equation02}) we have
\begin{eqnarray}
\sigma_{tot}(s)  =  4\pi   \int_{_{0}}^{^{\infty}}   b\,   db\,
[1-e^{-\chi_{_{R}}(s,b)}\cos \chi_{_{I}}(s,b)].
\label{eq01}
\end{eqnarray}
The parameter $\rho$, the real-to-imaginary ratio of the forward elastic scattering amplitude,
\begin{eqnarray}
\rho = \frac{\textnormal{Re}\,{\cal A}(s,t=0)}{\textnormal{Im}\,{\cal A}(s,t=0)},
\end{eqnarray}
can, in turn, be written using (\ref{equation02}) as
\begin{eqnarray}
\rho(s) = \frac{-\int_{_{0}}^{^{\infty}}   b\,  
db\, e^{-\chi_{_{R}}(s,b)}\sin \chi_{_{I}}(s,b)}{\int_{_{0}}^{^{\infty}}   b\,  
db\,[1-e^{-\chi_{_{R}}(s,b)}\cos \chi_{_{I}}(s,b)]}.
\label{eq03}
\end{eqnarray}

Since at high energies the soft and the semihard components of the scattering amplitude are closely related, we can assume that the eikonal
function is additive with respect to the soft and semihard (SH) parton interactions, and write the eikonal as
$\chi(s,b) = \chi_{_{soft}}(s,b) + \chi_{SH}(s,b)$ \cite{qcdmodel001,luna002,luna003}. We assume that the odd semihard eikonal is null and the
even contribution is obtained as follows. 
From considerations based on the QCD parton model and the unitarity condition (\ref{unitarity001}), the probability that neither nucleon is broken
up in a collision at impact parameter $b$ is given by $e^{-2\chi_{_{R}}(s,b)}$ \cite{qcdmodel001,qcdmodel002}. It follows that the even part of
the semihard eikonal contribution factorizes \cite{qcdmodel002,luna001,luna002,luna003},
\begin{eqnarray}
\chi^{+}_{SH}(s,b) = \frac{1}{2}\, \sigma_{_{QCD}}(s)\, W_{\!\!_{SH}}(s,b),
\end{eqnarray}
where $W_{\!\!_{SH}}(s,b)$ is an overlap density for the partons at $b$ and $s$,
\begin{eqnarray}
W_{\!\!_{SH}}(s,b) = \frac{1}{2\pi}\int_{0}^{\infty}dk_{\perp}\, k_{\perp}\, J_{0}(k_{\perp}b)\,G^{2}(s,k_{\perp}),
\label{overlap001}
\end{eqnarray}
and $\sigma_{_{QCD}}(s)$ is the usual QCD cross section 
\begin{eqnarray}
\sigma_{_{QCD}}(s) &=& \sum_{ij} \frac{1}{1+\delta_{ij}} \int_{0}^{1} dx_{1}
\int_{0}^{1} dx_{2} \int_{Q^{2}_{min}}^{\infty} d|\hat{t}|
\frac{d\hat{\sigma}_{ij}}{d|\hat{t}|}(\hat{s},\hat{t}) \nonumber \\
 & & \times f_{i/A}(x_{1},|\hat{t}|)f_{j/B}(x_{2},|\hat{t}|)\, \Theta \! \left( \frac{\hat{s}}{2} - |\hat{t}| \right),
\label{eq08}
\end{eqnarray}
where $\hat{s}$ and $\hat{t}$ are the Mandelstam invariants for the parton-parton collision, with $|\hat{t}|\equiv Q^{2}$ and
$x_{1}x_{2}s>2|\hat{t}|>2Q^{2}_{min}$, where $Q^{2}_{min}$ is the minimal momentum transfer in the semihard scattering
\cite{qcdmodel001,qcdmodel002,luna001,martin001}. In the above
expression $x_{1}$ and $x_{2}$ are the fractions of the momenta of the parent hadrons
$A$ and $B$ carried by the partons $i$ and $j$, with $i,j = q, \bar{q}, g$, $d\hat{\sigma}_{ij}/d|\hat{t}|$ is the differential cross
section for $ij$ scattering, and $f_{i/A}(x_{1},|\hat{t}|)$ ($f_{j/B}(x_{2},|\hat{t}|)$) is the parton $i$ ($j$)
distribution in the hadron $A$ ($B$).

Since the gluon distribution becomes very large as $x\to 0$, the parton-parton scattering processes used in the computation of $\chi_{SH}(s,b)$
must contain at least one gluon in the initial state. Thus, in the calculation of $\sigma_{_{QCD}}(s)$, we select the processes
$gg \to gg$, $qg\to qg$, $\bar{q}g\to \bar{q}g$, and  $gg\to \bar{q}q$. In fact, at $\sqrt{s}=7$ TeV and $Q_{min}=1.3$ GeV, their relative contribution
to $\sigma_{_{QCD}}(s)$ is around 98.8\% for the post-LHC fine-tuned parton distribution functions (PDFs) such as CT14 \cite{CT14}.

It is well known that these elementary processes are plagued by infrared divergences. Nevertheless, they can be regularized by considering an effective
charge whose finite infrared behavior is constrained by a gluon dynamical-mass scale
\cite{cornwall002}. The dynamical mass is intrinsically related to an infrared finite strong coupling $\bar{\alpha}_{s}$, and its existence, based on
the fact that the
nonperturbative dynamics of QCD may generate an effective momentum-dependent mass $M_g(Q^{2})$ for the gluons (while preserving the local $SU(3)_{C}$
invariance) \cite{aguilar001}, is strongly supported by QCD lattice results \cite{lattice001}. 
More specifically, lattice simulations reveal that the
gluon propagator is finite in the infrared region \cite{lattice002} and this behavior corresponds, from the Schwinger-Dyson formalism, to a massive gluon
\cite{cornwall001}.
In this Letter we adopt the functional forms of
$\bar{\alpha}_{s}$ and $M_g(Q^{2})$ obtained through the use of the pinch technique in order to derive a gauge invariant Schwinger-Dyson equation for the
gluon propagator and the triple gluon vertex \cite{cornwall002}:
\begin{eqnarray}
\bar{\alpha}_{s}(Q^{2})=\frac{4\pi}{\beta_{0}\ln\left[\left(Q^{2}+4M^{2}_{g}(Q^{2})\right)/\Lambda^{2}\right]},
\label{alfaS}
\end{eqnarray}
\begin{eqnarray}
M^{2}_{g}(Q^{2})=m^{2}_{g}\left[\frac{\ln\left[\left(Q^{2}+4m^{2}_{g}\right)/\Lambda^{2}\right]}{\ln\left(4m^{2}_{g}/\Lambda^{2}\right)}\right]^{-\frac{12}{11}},
\label{dgm}
\end{eqnarray}
where $\Lambda$ is the QCD scale parameter, $\beta_{0}=11-2n_{f}/3$ ($n_{f}$ is the number of flavors) and $m_{g}$ is the gluon mass scale. From lattice
simulations and phenomenological results its value is typically found to be of the order $m_{g} = 500 \pm 200$ MeV
\cite{luna001,luna002,luna003,lattice002,cornwall001}.
Note that in the limit $Q^{2} \to 0$ the effective charge $\bar{\alpha}_{s}(Q^{2})$ have an infrared fixed point, i.e. the dynamical mass tames the Landau
pole. 

We consider a semihard overlap density $W_{\!\!_{SH}}(s,b)$ that takes into account a ``broadening'' of the spatial distribution
\cite{luna002,lipari001},
\begin{eqnarray}
W_{\!\!_{SH}}(s,b;\nu_{\!\!_{SH}}) = \frac{\nu^{2}_{_{SH}}}{96\pi} (\nu_{_{SH}} b)^{3} K_{3}(\nu_{_{SH}} b),
\label{eq25}
\end{eqnarray}
where $\nu_{_{SH}}= \nu_{1}-\nu_{2}\ln (s/s_{0} )$, with $\sqrt{s_{0}}\equiv 5$ GeV, and $K_{3}(x)$ is the modified Bessel function of the second kind.
Here, $\nu_{1}$ and $\nu_{2}$ are constants to be fitted. The above expression is obtained assuming a dipole form factor
$G(s,k_{\perp};\nu_{\!\!_{SH}}) = \left(\nu^{2}_{\!\!_{SH}} / (k^{2}_{\perp} + \nu^{2}_{\!\!_{SH}}) \right)^{2}$ in equation (\ref{overlap001}). 
The energy dependence of $W_{\!\!_{SH}}(s,b)$ is suggestive of a parton picture where the semihard interactions are dominated by gluons whereas
soft interactions are mainly related to interactions among
valence quarks \cite{luna002}. Hence even and odd soft
contributions based on Regge-Gribov formalism \cite{barone001,forshaw001,kmr002,kmr003} are very well suited for our purposes:
\begin{eqnarray}
\chi_{_{soft}}^{+}(\tilde{s},b) = \frac{1}{2}\,W_{_{soft}}(b;\mu^{+}_{_{soft}}) 
\left[ A  +  \frac{B}{\sqrt{\tilde{s}}}  + C \tilde{s}^{\lambda} \right],
\label{chDGM.10}
\end{eqnarray}
\begin{eqnarray}
\chi_{_{soft}}^{-}(\tilde{s},b) = \frac{1}{2}\, W_{_{soft}}(b;\mu^{-}_{_{soft}})\, \frac{D}{\sqrt{\tilde{s}}}\, e^{-i\pi/4},
\label{chDGM.32}
\end{eqnarray}
where $\tilde{s}\equiv s/s_{0}$, $\mu^{-}_{_{soft}}\equiv 0.5$ GeV, $\lambda = 0.12$, and $A$, $B$, $C$, $D$ and $\mu^{+}_{_{soft}}$ are fitting parameters.
The term $C \tilde{s}^{\lambda}$ in (\ref{chDGM.10}) represents the contribution of a soft Pomeron, whose behavior is well described by a power
close to $\tilde{s}^{\, 0.12}$ \cite{kmr003}. In order to study the soft contribution to the total cross sections at high energies we have also attempted
a Froissaron \cite{odderon002} term $C\ln^{2}(\tilde{s})$ in the place of the power, but we have not observed any significant difference between the
two cases. The soft form factors are simply ``static'' versions of (\ref{eq25}),
\begin{eqnarray}
W_{_{soft}}(b;\mu^{+,-}_{_{soft}}) = \frac{(\mu^{+,-}_{_{soft}})^{2}}{96\pi} (\mu^{+,-}_{_{soft}} b)^{3} K_{3}(\mu^{+,-}_{_{soft}} b) .
\label{chDGM.19}
\end{eqnarray}
In this scenario the soft odd eikonal $\chi^{-}_{_{soft}}(\tilde{s},b)$ is significant only in the low-energy regime.
We ensure the correct analyticity properties of the amplitudes from the substitution $s\to se^{-i\pi/2}$ throughout Eqs.
(\ref{eq08}), (\ref{eq25}) and (\ref{chDGM.10}) \cite{luna001,martin001}, valid for an even amplitude and equivalent to the use of a dispersion relation.
The Eq. (\ref{chDGM.32}) already have the correct analyticity structure. The prescription $s\to se^{-i\pi/2}$ is applied on Eq. (\ref{eq08}) taking into
account the fact that $\sigma_{_{QCD}}(s)$ corresponds to the real part of a more general complex analytic function $F(-is)$. Thus $\sigma_{_{QCD}}(s)$
can be correctly reproduced by means of a novel flexible function $F$, and if $F$ is a holomorphic function in a region $\Omega$, the analytical
continuation ensures the uniqueness of the real part of $F(-is)$. Specifically, we consider a complex analytic parametrization given by
\begin{eqnarray}
  F(-is) &=& b_{1} + b_{2}\,e^{b_{3}[X(s)]^{1.01\, b_{4}}} + b_{5}\,e^{b_{6}[X(s)]^{1.05\, b_{7}}} \nonumber \\
  & & + b_{8}\,e^{b_{9}[X(s)]^{1.09\, b_{10}}} ,
\label{eq:sigQCD_fit}
\end{eqnarray}
where $b_{1},...,b_{10}$ are free fit parameters and $X(s)=\ln\,\ln(-is)$. In this procedure the $b_{1},...,b_{10}$ parameters are then adjusted to
optimally satisfy the relation $\textnormal{Re}F(-is)=\sigma_{_{QCD}}(s)$ with less than 1$\%$ error. It follows that
$\textnormal{Im}F(-is)=\textnormal{Im}\sigma_{_{QCD}}(s)$. In the case of CT14, the data reduction yelds: $b_{1} = $ 100.220   GeV$^{-2}$,
$b_{2} =$ 0.434 $\times$ 10$^{-1}$ GeV$^{-2}$,
$b_{3} = $ 1.274,
$b_{4} = $ 1.919,
$b_{5} = $ 0.122 $\times$ 10$^{-7}$ GeV$^{-2}$,
$b_{6} = $ 14.050, 
$b_{7} = $ 0.504, 
$b_{8} = $ 3.699 $\times$ 10$^{3}$ GeV$^{-2}$,
$b_{9} = $ -80.280,  
$b_{10} = $ -2.632.

\section{Results and Conclusions} 

In our analyses we carry out a global fit to forward ($t=0$) scattering
data from $\sqrt{s_{min}} = 10$ GeV to LHC energies.
We use data sets compiled and analyzed by the Particle Data Group \cite{PDG} as well as the recent data at LHC from the ATLAS \cite{atlas001} and the 
TOTEM \cite{antchev001,antchev002} Collaborations, with the statistic and systematic errors added in quadrature.
Specifically, we fit to the total cross sections, $\sigma_{tot}^{pp}$ and $\sigma_{tot}^{\bar{p}p}$,
and the ratios of the real to imaginary part of the forward scattering amplitude, $\rho^{pp}$ and $\rho^{\bar{p}p}$. 
In doing this we utilize a $\chi^{2}$ fitting procedure, adopting an interval $\chi^{2}-\chi^{2}_{min}$ corresponding to the projection of the
$\chi^{2}$ hypersurface containing 68.3\% of probability, namely 1 $\sigma$. The value of $\chi^{2}_{min}$ is distributed as a $\chi^{2}$ distribution
with $\zeta$ degrees of freedom. As a convergence criteria we consider only data reductions which
imply positive-definite covariance matrices, since theoretically the covariance matrix for a physically motivated function must be, at the minimum,
positive-definite. As tests of goodness-of-fit we adopt the chi-square per degree of freedom $\chi^{2}/\zeta$ and the integrated probability
$P(\chi^{2};\zeta)$.

\begin{table}
\centering
\caption{Best fit parameters of the QCD-based model with CT14 PDFs. The dynamical gluon mass scale was set to $m_{g}=400$ MeV, while the minimal 
momentum transfer was set to $Q_{min} = 1.3$ GeV.}
\begin{ruledtabular}
\begin{tabular}{cc}
$\nu_{1}$ [GeV] & 1.70$\pm$0.21 \\
$\nu_{2}$ [GeV] & 0.015$\pm$0.014 \\
$A$ [GeV$^{-2}$] & 88.2$\pm$9.6 \\
$B$ [GeV$^{-2}$] & 51.7$\pm$9.7 \\
$C$ [GeV$^{-2}$] & 27.2$\pm$5.7 \\
$D$ [GeV$^{-2}$] & 24.2$\pm$1.4 \\
$\mu^{+}_{_{soft}}$  [GeV] & 0.90$\pm$0.16 \\[0.5ex]
\hline \\[-2.2ex]
$\zeta$ & 166 \\
$\chi^{2}\!/\zeta$ & 1.30 \\
$P(\chi^{2};\zeta)$ & 5.4\,$\times$\,$10^{-3}$ \\
\end{tabular}
\end{ruledtabular}
\end{table}

In performing calculations using the QCD-based formalism it is necessary to make some choice out of the many PDFs
available. We have chosen the CT14 PDFs from a global analysis by the CTEQ-TEA group \cite{CT14}. The CT14 is a next generation of PDFs that
include data from the LHC for the first time. The CT14 set, that also includes updated data from the HERA and Tevatron experiments, is available
from LHAPDF \cite{lhapdf}. The CTEQ-TEA team have incorporated important enhancements that made CT14 instrumental to our analysis. More
specifically, CT14 employs, at initial scale $Q_{0}$, a flexible parametrization based on the use of Bernstein polynomials, which permit a better
capture of the PDFs variations in the DGLAP evolution; CT14 also includes for the first time measurements of inclusive production of
vector bosons \cite{inclusiveboson}, as well as of jets \cite{jetsinclusion}, at 7 and 8 TeV as input for the global fits. The inclusion of
single-inclusive jet production measurements at 7 TeV is sufficient to constrain the gluon distribution function and leads to a clear improvement in
the gluon PDF uncertainty.

The values of the fitted parameters are given in Table 1. The $\chi^{2}/\zeta$ for the fit was obtained for 166 degrees of freedom.
The results of the fits to $\sigma_{tot}$ and $\rho$ for both $pp$ and $\bar{p}p$ channels are displayed, together with the
experimental data, in Figs. 1 and 2. In Fig. 1 we include, only as illustration, two estimates of $\sigma_{tot}$ from cosmic-ray
experiments: the Auger result at 57 TeV \cite{auger001} and the Telescope Array result at 95 TeV \cite{ta001}.

\begin{figure}
\label{figpaperluna01}
\begin{center}
\includegraphics[height=.25\textheight]{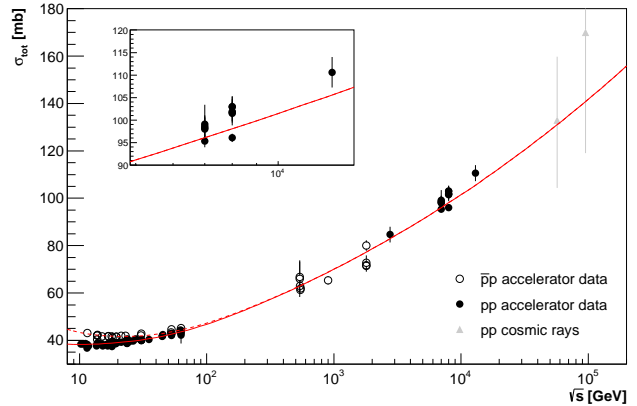}
\caption{Total cross section for $pp$ (solid curve) and $\bar{p}p$ (dashed curve) channels.}
\end{center}
\end{figure}

\begin{figure}
\label{figpaperluna02}
\begin{center}
\includegraphics[height=.25\textheight]{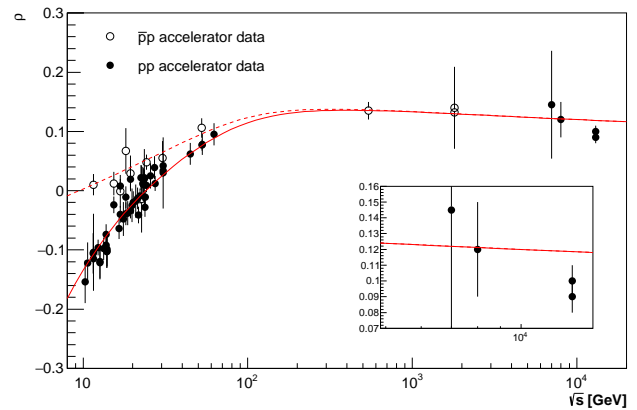}
\caption{Ratio of the real to imaginary part of the forward scattering amplitude for $pp$ (solid curve) and $\bar{p}p$ (dashed curve) channels.}
\end{center}
\end{figure}

Notice that our QCD-based model allows us to describe in a satisfactory way the forward scattering observables
$\sigma_{tot}^{pp}$ and $\rho^{pp}$ from $\sqrt{s}=10$ GeV to 13 TeV. Our predictions for the total cross section and $\rho$ parameter at
$\sqrt{s}=13$ TeV are $\sigma^{pp}_{tot} = 105.6^{+7.9}_{-6.3}$ mb and $\rho^{pp} = 0.1182^{+0.0074}_{-0.0057}$, respectively. The uncertainties
in our theoretical predictions have been estimated taking into account the typical uncertainty of the gluon mass scale and the PDFs uncertainties
on the production cross sections at the LHC, as discussed in Ref. \cite{luna002}.
These results show that our model, whose formulation is
compatible with analyticity and unitarity constraints, is well suited for predictions of forward observables to be measured at high-energy scales.
We argue that the distinction in the model between semihard gluons, which participate in hard parton-parton scattering, and soft gluons, emitted in
any given parton-parton QCD radiation process, is highly relevant to understanding our simultaneous description of $\sigma_{tot}$ and
$\rho$ data over a wide range of energies. The introduction of infrared properties of QCD, by considering that the nonperturbative dynamics of Quantum
Chromodynamics generate an effective gluon mass, is of central importance since at high energies the soft and the semihard components of the scattering
amplitude are closely related \cite{ryskin001}. Most importantly, from a rigorous statistical point of view, our analysis shows that the TOTEM measurements
can be simultaneously described by a QCD scattering amplitude dominated by only single crossing-even elastic terms. Crossing-even dominance
entails that $\sigma^{\bar{p}p}_{tot}-\sigma^{pp}_{tot}\to 0$ and $\rho^{\bar{p}p}_{tot}-\rho^{pp}_{tot}\to 0$ as $s \to \infty$.

However, it is worth noting that the values of $\rho$ measured by TOTEM at $\sqrt{s}=13$ TeV are lower than our prediction: $\rho = 0.09 \pm 0.01$ and
$\rho = 0.10 \pm 0.01$ \cite{antchev002}. Note that these experimental values are measured with quite extraordinary precision, and even taking into
account our theoretical uncertainties the model prediction is not compatible with the $\rho$ values ​​measured by TOTEM at $\sqrt{s}=13$ TeV.
These measurements, if eventually confirmed by other experimental groups, suggest that a crossing-odd elastic term may play a central role in the soft
and semihard interactions at high energies. In other words, the TOTEM $\rho$ data suggest the existence of an Odderon.

\section*{Acknowledgments}

This research was partially supported by the Conselho Nacional de
Desenvolvimento Cient\'{\i}fico e Tecnol\'ogico (CNPq) under the grant 141496/2015-0. 
DAF acknowledges support by the project INCT-FNA (464898/2014-5).
EGSL acknowledges the financial support from the ANII-FCE-126412 project.


\end{document}